\documentclass[fleqn]{vch-book}
\usepackage{amsmath}
\usepackage{amssymb}
\usepackage{makeidx}
    \makeindex
\usepackage{times}
\usepackage{tabularx}
\usepackage{booktabs}
\usepackage{epsfig}

\begin{document}
\chapter{ Nonlinear Vibration in Gear Systems
\tocauthor{G. Litak and M.I. 
Friswell}}\label{chap:editbooks}
\authorafterheading{Grzegorz Litak$^1$ and Michael I. Friswell$^2$} 

\affil{ $^1$ Department of Applied Mechanics, \\ Technical University of 
Lublin,
Nadbystrzycka 36, PL-20-618 Lublin, \\ Poland \\ \\
$^2$ Department of Areospace Engineering,  \\ 
University of Bristol,
Queens Building,  Bristol BS8 1TR,\\ United Kingdom}

Gear box dynamics is characterised by a periodically
changing
stiffness and a backlash which  can lead to a
 loss of the contact between the teeth. Due to backlash, the gear system
has piecewise linear stiffness characteristics and, in consequence,  
can vibrate
regularly or chaoticaly depending on the system parameters and the initial conditions.
We examine the possibility of a  nonfeedback system  control by introducing a weak 
resonant excitation term   and through 
adding an additional degree
of freedom to account for shaft flexibility on one side of the gearbox.
We shall show that by correctly choosing
the coupling values the system vibrations may be controlled.

\section{Introduction}
Gear box dynamics is based on  a periodically changing 
meshing stiffness complemented  by  a nonlinear  effect of  backlash 
between the teeth. 
Recently, their
regular and chaotic vibrations have been predicted theoretically and examined 
experimentally  
\cite{Kah90,Sat91,Bla95,Kah97,Rag99,War00,Lit03}. 
The theoretical description of this phenomenon
has been based mainly on single degree of freedom models 
\cite{Kah90,Sat91,Bla95,Kah97,Rag99,War00,Sza96} 
or multi degree models neglecting
backlash \cite{Sch86,War00b}. This paper examines the possibility of 
taming chaotic vibrations and reducing the amplitude by nonfeedback 
control methods. Firstly, we will introduce a weak resonant excitation 
through an additional small external excitation term (torque)  with a different 
phase \cite{Cha01,Cao03}. 
Secondly, we will examine  the effect of  an 
additional degree of freedom to account for
shaft flexibility on one side of the gearbox \cite{Lit03}.
 This may also be regarded as a single mode approximation to the
torsional system dynamics, or alternatively as a 
simple model for a vibration neutraliser or absorber
installed in one of the gears.

\section{One Stage Gear Model}
\begin{figure} 
\vspace{1cm} \begin{center}
\includegraphics[scale=0.5]{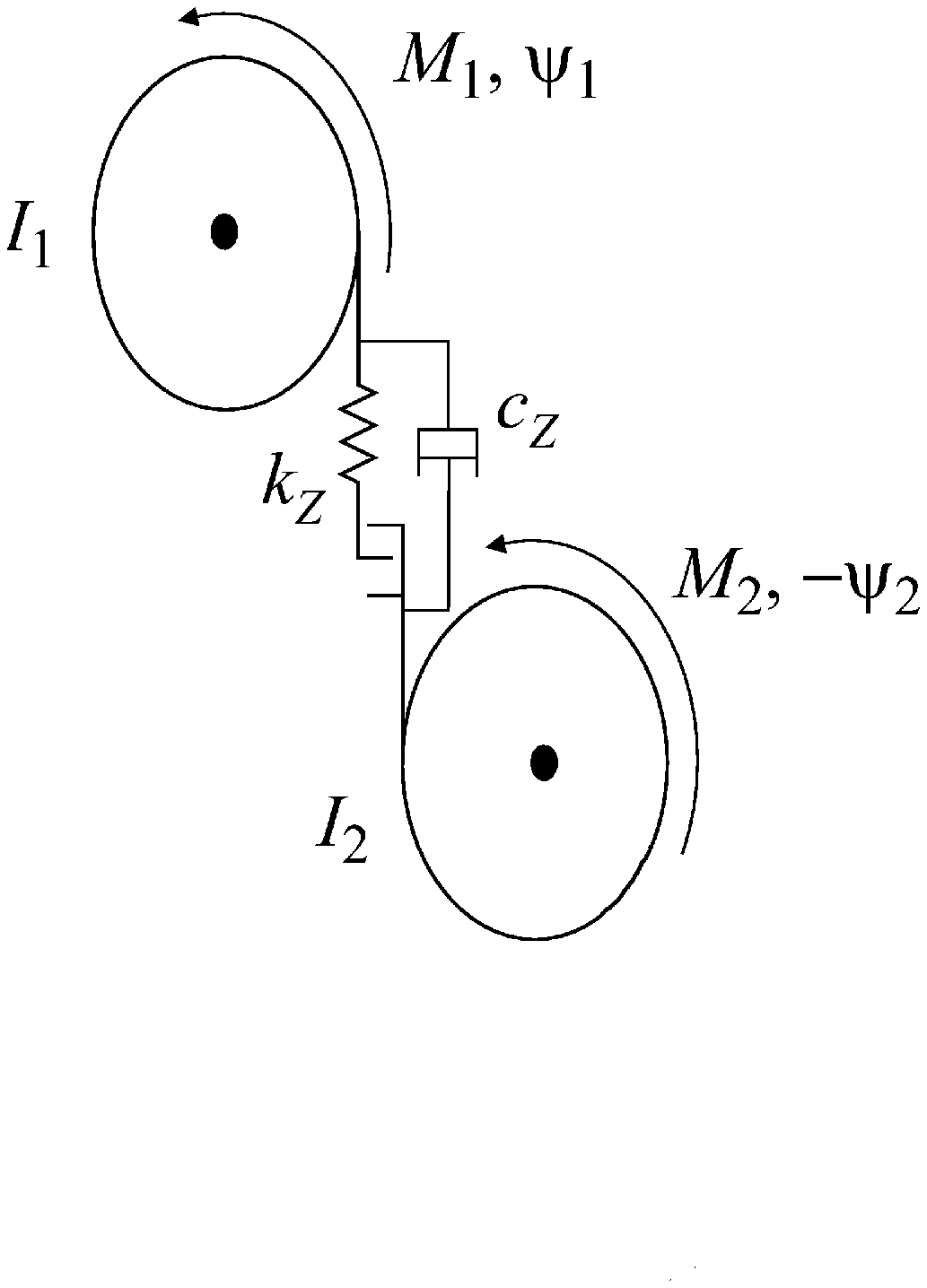}  
\vspace{-1cm} 
\caption{ \label{GLf1}
The physical model of a one stage transmission gear system.}
\end{center}
\end{figure}

We start our analysis with modelling the relative vibrations of a single stage transmission
gear \cite{War00}. 
Figure \ref{GLf1}  shows a schematic picture of the physical system. 
The gear wheels are shown with
moments of inertia $I_1$ and $I_2$ and are coupled by the stiffness and damping of the teeth
mesh, represented by $k_Z$ and $c_Z$.

The equations of motion of the system may be written in terms of the
two degrees of freedom, $\psi_1$, $\psi_2$ that represent the rotational angles
 of the gear wheels. These angles are those that remain
after the steady rotation of the system is removed \cite{War00}.
Thus, if the backlash is initially neglected, we have

\begin{eqnarray}
I_1 \ddot{\psi}_1 + \left[ k_Z(r_1\psi_1-r_2\psi_2)+
c_Z(r_1\dot{\psi}_1-r_2 \dot{\psi}_2) \right] r_1 & = & M_1,
 \nonumber \\
I_2 \ddot{\psi}_2 - \left[ k_Z(r_1\psi_1-r_2\psi_2)+
c_Z(r_1\dot{\psi}_1-r_2 \dot{\psi}_2) \right] r_2 & = & -M_2, \label{GLe1}
\end{eqnarray}
where $r_1$ and $r_2$ are the radii of the gear wheels and the overdot represents
differentiation with respect to time $t$.

\begin{figure}
\vspace{1cm} \begin{center}
\includegraphics[scale=0.5,angle=-90]{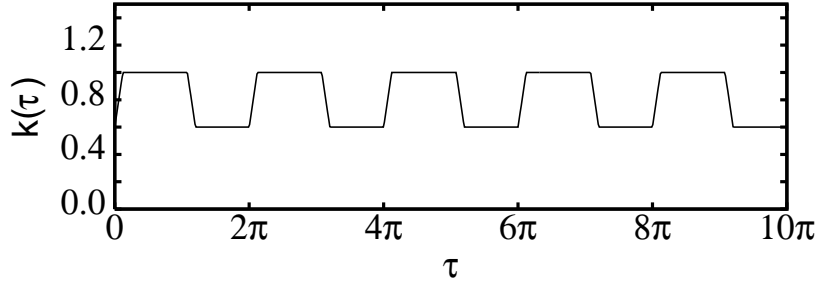} 
\caption{\label{GLf2} The meshing stiffness $k(\tau)$.}
\end{center}
\vspace{0cm}
\end{figure}

\begin{figure}
\vspace{1cm} \begin{center}
\includegraphics[scale=0.5,angle=-90]{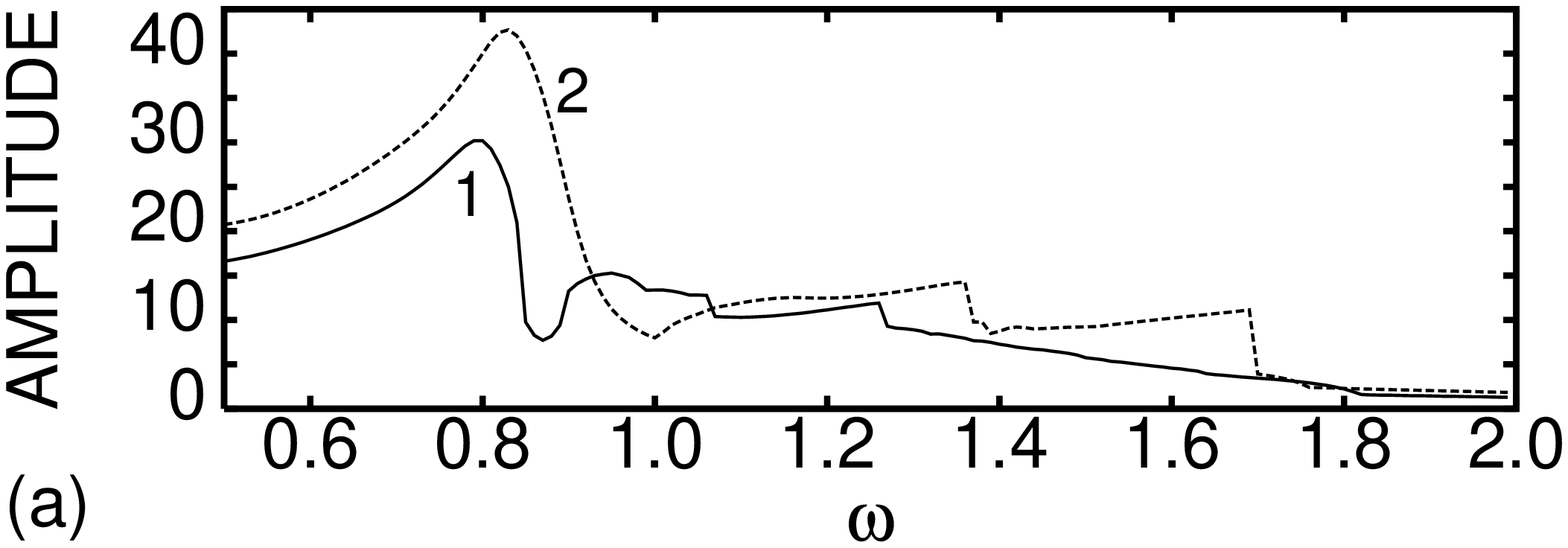}
\includegraphics[scale=0.5,angle=-90]{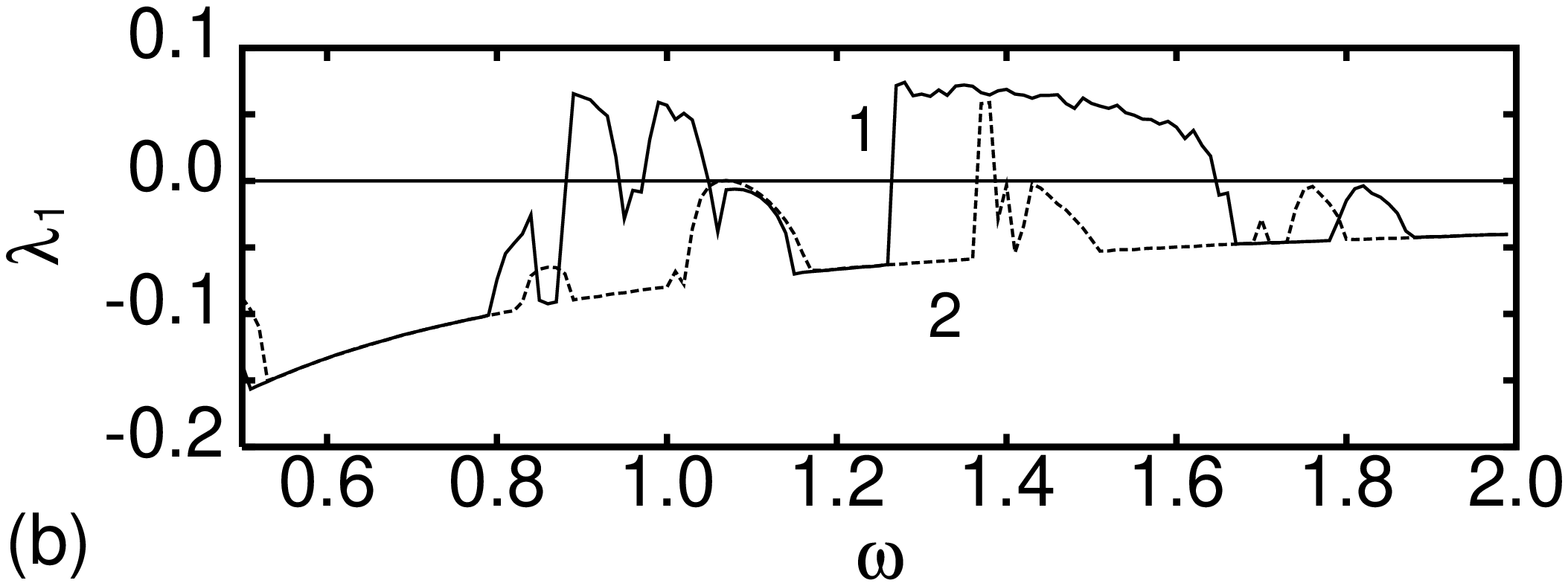}
\caption{ \label{GLf3}  The amplitude of vibration (a) and the maximal Lyapunov exponent $\lambda_1$ (b) versus frequency 
$\omega$ 
obtained by simulation.  The initial conditions were assumed to be   $x_0=-2.0$ and $v_0=\dot{x}_0=-0.5$ 
for small $\omega$ ($\omega_0=0.1$) and for each new  $\omega$ ($\omega_{i+1}$) calculations were 
performed for  
$400$ 
excitation cycles. Each time  
new initial conditions $(x_0,v_0)$ were chosen as the last pair of values of $(x,v)$ for previous 
$\omega_{i}$ ($(x_0,v_0)|_{new}=(x,v)|_{old}$). Curves '1' and '2' correspond to cases with $D=0$ and $D=2$, respectively.}
\end{center}
\vspace{0cm} 
\end{figure}

\begin{figure}
\vspace{1cm}
\hspace{-0.7cm}
\includegraphics[scale=0.4,angle=-90]{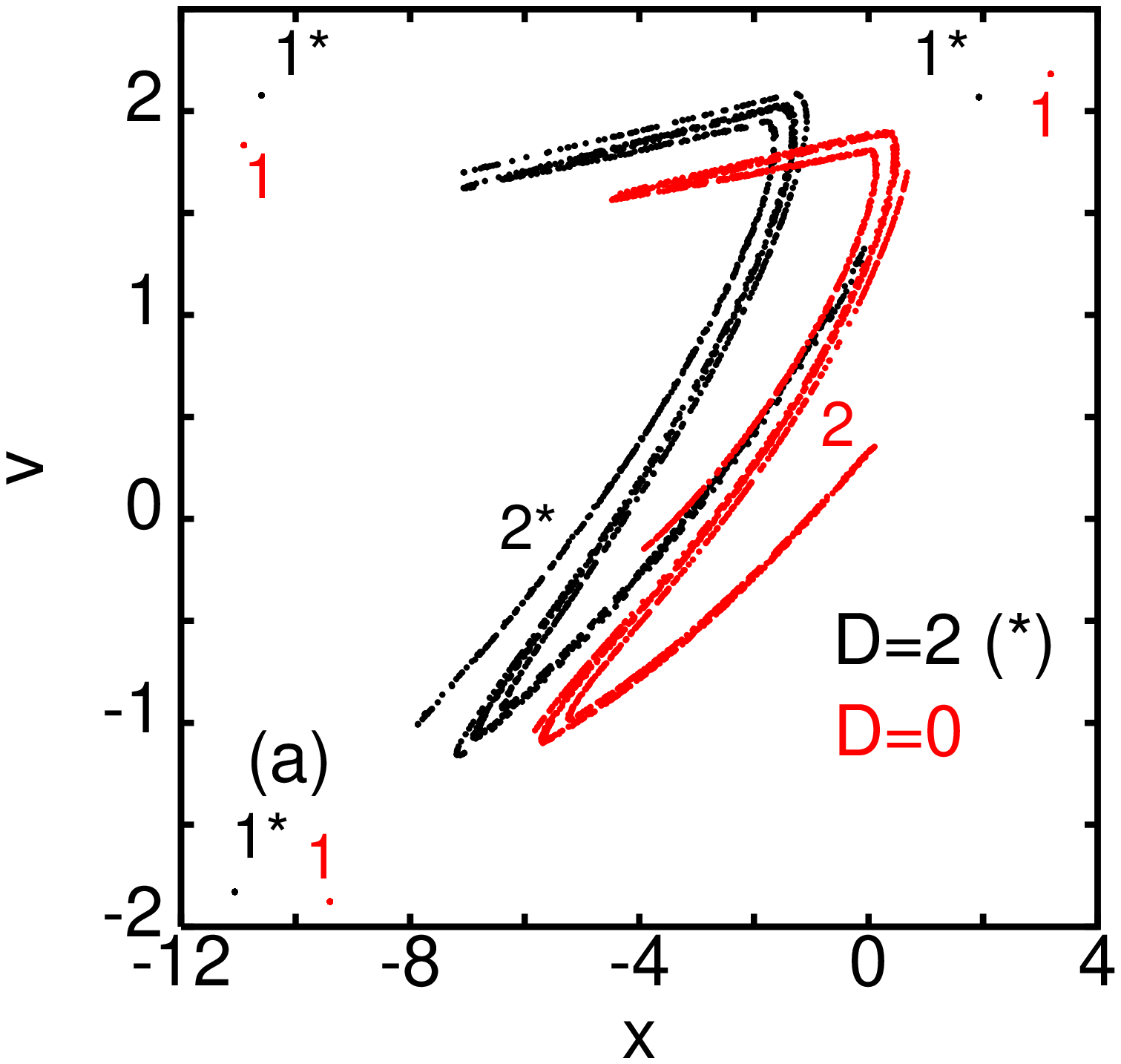} \hspace{-1.2cm} 
\includegraphics[scale=0.4,angle=-90]{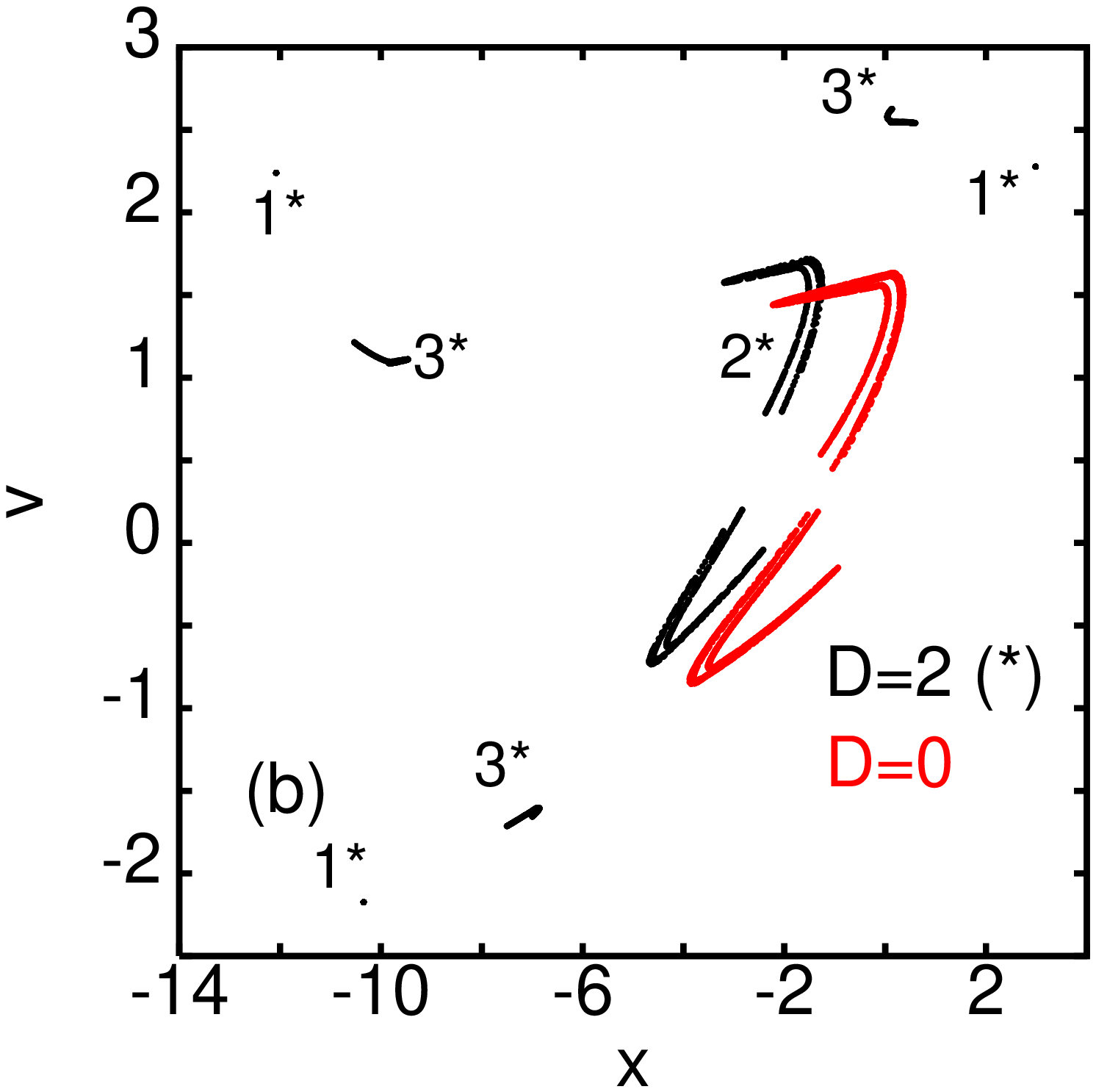} 
\caption{\label{GLf4} The Poincare maps calculated for ten randomly chosen initial conditions for $\omega=1.5$ (a) and 
$\omega=1.6$  the gray colour corresponds to simulations with respect to (Eq. \ref{GLe2}) while the black 
one to (Eq. \ref{GLe6}), respectively.}
\vspace{0cm}
\end{figure}

 In fact, it is possible to reduce the above two equations  to one using the new relative
displacement coordinate  $x=r_1\psi_1 - r_2 \psi_2$. That coordinate
represents the relative displacement of the gear wheels at the teeth. 
The equation of motion for $x$ is obtained 
by subtracting $r_2/I_2$ times first equation from $r_1/I_1$ 
times the second one (Eqs. \ref{GLe1}). Thus, in non-dimensional form, the  
equation of motion can be rewritten as
\begin{equation}
\frac{{\rm d}^2}{{\rm d} \tau^2}x + \frac{2 \zeta}{\omega} \frac{{\rm d}}{{\rm d} \tau}x+ \frac{k(\tau) 
g(x,\eta)}{\omega^2}   =  \frac{\bar{B}(\tau)}{\omega^2} 
= \frac{B_0 + B_1 cos (\tau+ \Theta )}{\omega^2},
\label{GLe2}
\end{equation}
where the parameters are easily derived from Eqs. (\ref{GLe1}-\ref{GLe2}), as
\begin{eqnarray}
        \tau &=& \omega t, \nonumber \\
	2 \zeta & = & c_Z \left[ r_1^2 /I_1 + r_2^2 / I_2 \right], 
 \label{GLe3}
 \\
  \bar{B}(\tau) & = & r_1M_1/I_1 + r_2M_2/I_2. \nonumber
\end{eqnarray}
$\zeta$, $k(\tau)$, $g(x,\eta)$ and $\bar{B}(\tau)$ are defined as  in reference \cite{War00}. 
Note that the backlash and time dependent meshing stiffness have been included by 
rewriting the meshing force as \cite{War00}
\begin{equation}
	k_Z \left[ r_1^2 /I_1 + r_2^2 / I_2 \right] x = k(\tau) g(x,\eta).
	\label{GLe4}
\end{equation}
Equations (\ref{GLe3}) and (\ref{GLe4}) have assumed that the moments on the gear system are
composed of a sinusoidal moment at frequency $\omega$ with a constant offset.

The meshing stiffness is periodic and the backlash is described by a piecewise
linear function. Figure \ref{GLf2} shows a typical time dependent mesh stiffness 
variation $k(\tau)$ \cite{War00} and the backlash 
is modelled for a clearance $\eta$ as

\begin{equation}
g\left( {x,\eta } \right) = \left\{ {\begin{array}{*{20}c}
   x & \hspace{8mm} & {x \ge 0}  \\
   0 & & { - \eta  < x < 0}  \\
   {x + \eta } & & {x \le \eta }.  \\
\end{array}} \right.
\label{GLe5}
\end{equation}

\section{Vibrations of a Gear System in Presence of a Weak Resonance Term}
Now we are going to examine the gear system described by Eqs. (\ref{GLe1}-\ref{GLe5}) subjected to 
to an additional external periodic excitation with a relatively small value $D$  
\begin{equation}
\frac{{\rm d}^2}{{\rm d} \tau^2}x + \frac{2 \zeta}{\omega} \frac{{\rm d}}{{\rm d} \tau}x+ \frac{k(\tau)
g(x,\eta)}{\omega^2}   = 
\frac{B_0 + B_1 cos (\tau+ \Theta ) + D cos (\tau+ \Theta' )}{\omega^2},
\label{GLe6}
\end{equation}
where the phase angle $\Theta' \neq \Theta$. Such inclusion has been discussed earlier in the literature
as a suitable method of
chaos control for Josephson junction circuits \cite{Cha01} and Froude pendulum motion
\cite{Cao03}.  In those papers the authors claimed that the resonant term, if used adequately, can tame or 
include chaotic motion. 
We have performed numerical simulations of Eqs. (\ref{GLe2}) and (\ref{GLe6})  to highlight the effect of 
the addition of resonant term $D$ on the basic equation of motion (Eq. \ref{GLe2}). In particular we were 
interested in changes in vibration amplitude. The type of 
motion was analysed using Lyapunov exponents obtained by the algorithm of Wolf {\em et al.}  
\cite{Wol88}.  
System parameters have been used as in \cite{War00}:
 $\omega=1.5$, $B_0=1$, $B_1=4$, $\Theta=0$,  $\zeta = 0.08$, $\eta = 10$  
while the additional excitation term \ref{GLe6} was 
introduced as $D=2$,
 $\Theta'=0.1$. 

Figure \ref{GLf3}a shows 
the calculated vibration amplitude of the relative motions of the gear wheels, $A$,  defined as
\begin{equation}
A=\left| \frac{x_{max}-x_{min}}{2} \right|.
\label{GLe7}
\end{equation}
The maximal Lyapunov exponent 
$\lambda_1$ versus frequency
$\omega$,
obtained by simulation, is plotted in Fig. \ref{GLf3}b.  Note, the initial conditions were assumed to be   
$x_0=-2.0$ and 
$v_0=\dot{x}_0=-0.5$
for small starting $\omega$ ($\omega=0.1$) and for each new  $\omega_{i+1}$ calculations were performed 
for  
$400$
excitation cycles with
where new initial conditions were the last pair of values of $(x,v)$ for the previous frequency $\omega_{i}$. 
Curves marked by '1' correspond to solutions of Eq. \ref{GLe2} while those marked by '2' correspond to
Eq. \ref{GLe6}. 
One can easily note that the amplitude of the modified system is slightly larger but simultaneously the 
system behaves
more regularly.  In most of cases where the original system was in a chaotic state ($\lambda_1 > 1$ for the 
curve '1', Fig. \ref{GLf3}b) it vibrates  regularly  in the presence resonant term  ($\lambda_1 < 1$ for the 
curve 
'2' in Fig. \ref{GLf3}b). Unfortunately the taming of the chaotic motion was accompanied by 
additional features 
visible in Fig. \ref{GLf3}a. Clearly, the curve '2' describing the amplitude of motion shows a number of 
jumps, i.e. for $\omega \approx 1.36$ and 1.70. These jumps may be associated with transitions to other 
solutions (attractors)
 with changing $\omega$. This effect is typical for many nonlinear systems but it seems to be more 
transparent for $D \neq 0$.  
To explore further this effect we have also calculated Poincare 
maps for many initial conditions chosen randomly.  The results for frequencies $\omega=1.5$ and 
1.6 are shown in Fig.  \ref{GLf4}. In this figure, one can see that for $\omega=1.5$ attractors 
(in gray colour) for $D=0$ and (in black colour) for $D \neq 0$ are similar. Here, numbers '1' and '2' 
denote the regular and 
chaotic attractors, respectively. 
But if we move $\omega$ to a larger value  ($\omega$= 1.6) the regular attractor for $D=0$ 
disappears. The attractor remains for $D \neq 0$, but interestingly another chaotic
 attractor emerges. For some other 
frequencies additional attractors also exist, which supports the proposed explanation of the jumping 
phenomenon.

\section{Vibrations of a Gear System with a Flexible Shaft}

\begin{figure}
\vspace{0.5cm} \begin{center}
\includegraphics[scale=0.5]{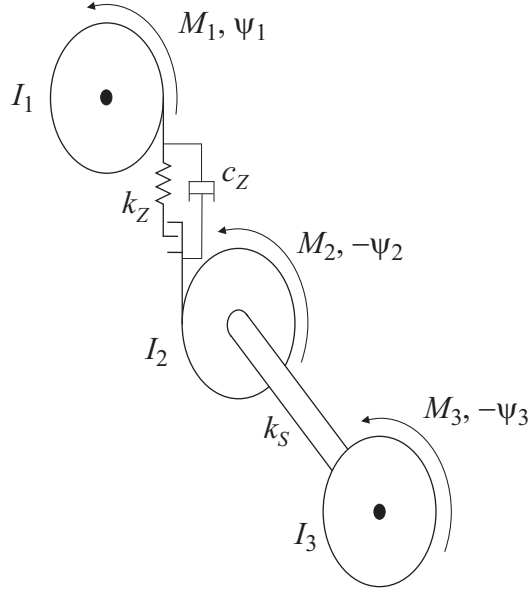}
\vspace{1cm}
\caption{\label{GLf5}
 The physical model of the gear model with an additional degree of freedom
introduced by  a  disk fixed by a flexible shaft to one of gear wheels.
}
\end{center}
\end{figure}

Here we examine the effect of adding an
additional degree of freedom to the gear system (Fig. \ref{GLf5}) by using
a flexible shaft on one side of the gearbox \cite{Lit03}. 

Now the equations of motion of the system may be written in terms of the
three degrees of freedom, $\psi_1$, $\psi_2$ and $\psi_3$, related to the rotational angles 
of the gear wheels and disk, and have the following form

\begin{eqnarray}
I_1 \ddot{\psi}_1 &+& \left[ k_Z(r_1\psi_1-r_2\psi_2)+
c_Z(r_1\dot{\psi}_1-r_2 \dot{\psi}_2) \right] r_1  =  M_1,
\nonumber \\
I_2 \ddot{\psi}_2 &-& \left[ k_Z(r_1\psi_1-r_2\psi_2)+
c_Z(r_1\dot{\psi}_1-r_2 \dot{\psi}_2) \right] r_2 - k_S ( \psi_3-\psi_2 )  =  -M_2, \nonumber
\\
I_3 \ddot{\psi}_3 &+& k_S ( \psi_3-\psi_2 )  =  -M_3.
\label{GLe8}
\end{eqnarray}

\begin{figure}
\vspace{1cm} \begin{center}
\includegraphics[scale=0.5,angle=-90]{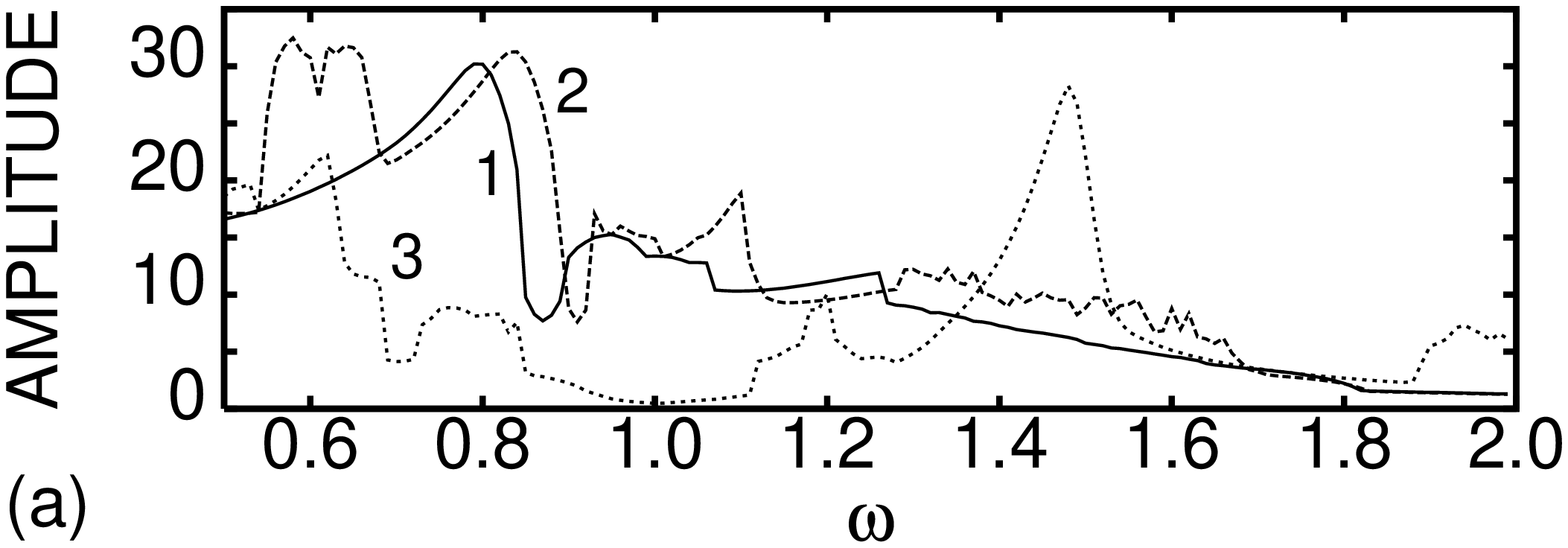} 
\includegraphics[scale=0.5,angle=-90]{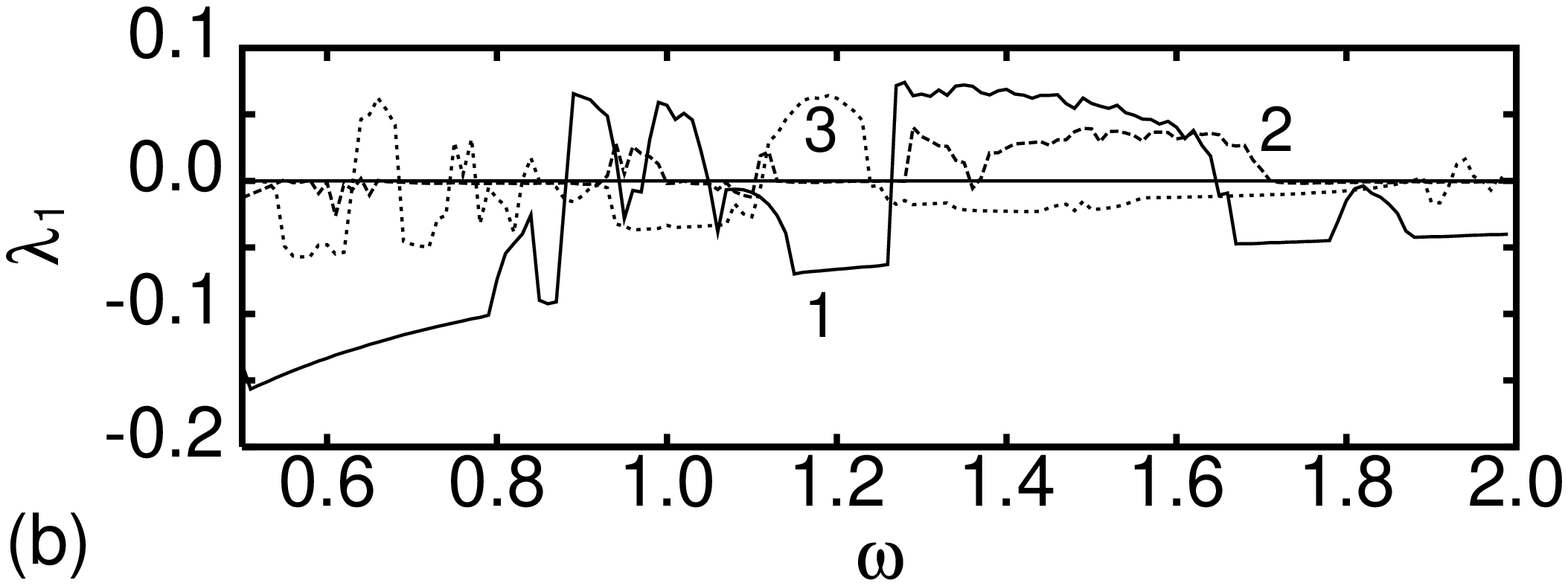}
\includegraphics[scale=0.5,angle=-90]{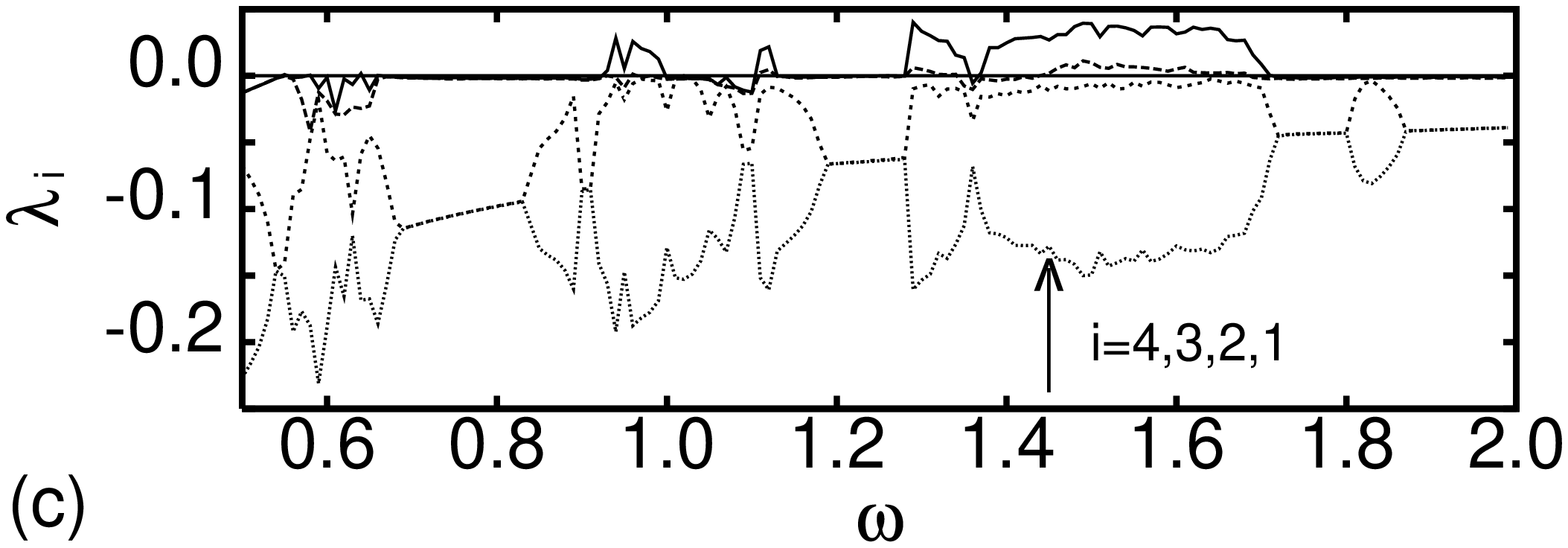}
\includegraphics[scale=0.5,angle=-90]{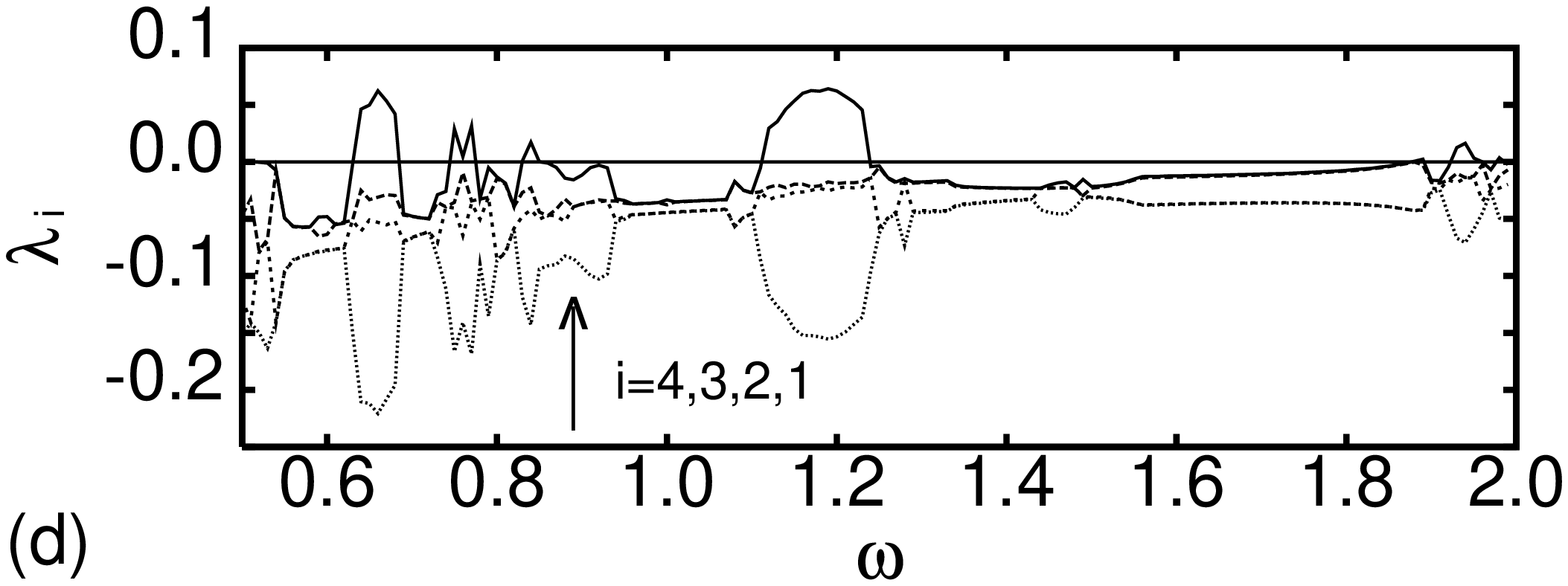}
\caption{\label{GLf6}
 The amplitude of vibration (a) and the maximal Lyapunov exponent $\lambda_1$ (b) versus frequency
$\omega$
obtained by simulations.  The initial conditions were assumed to be   $x_{10}=-2.0$, 
$v_{10}=\dot{x}_{10}=-0.5$ and $x_{20}=0$ $v_{20}=\dot{x}_{20}=0$
for small $\omega$ ($\omega=0.1$) and for each new  $\omega_{i+1}$ calculations were performed for  $400$
excitation cycles 
where the new initial conditions were the last pair of values of $(x_1,v_1,x_2,v_2)$ for the previous 
$\omega_{i}$.
Curves '1','2','3' were obtained for $k_S=0$, 0.1 and 1.0, respectively.
Four nonzero Lyapunov exponents $\lambda_i$ ($i=1,2,3,4$) versus frequency $\omega$ are shown
for $k_S=0.1$ (c) and $k_S=1.0$ (d).
}
\end{center}
\vspace{0cm}
\end{figure}

As in the previous case (Eqs. \ref{GLe1}, \ref{GLe2}) the above set of equations may be decoupled
by using the new relative 
displacement coordinates  $x_1=r_1\psi_1 - r_2 \psi_2$ and $x_2=r_2(\psi_3 -
\psi_2)$. Thus we 
reduce the number of equations from three to two

\begin{eqnarray}
\frac{{\rm d}^2}{{\rm d} \tau^2}x_1  &+& \frac{2 \zeta}{\omega} \frac{{\rm d}}{{\rm d} \tau}x_1 + 
\frac{k(\tau) 
g(x_1,\eta)}{\omega^2} + \frac{\beta_1 k_S x_2}{\omega^2} \nonumber \\ &=&  \frac{B_0 + B_1 cos ( \omega t+ \Theta 
)}{\omega^2},
\label{GLe9} \\
\frac{{\rm d}^2}{{\rm d} \tau^2}x_2   &+& \frac{\beta_2 k_S x_2}{\omega^2} + \frac{2 \beta_3 
\zeta}{\omega} 
\frac{{\rm d}}{{\rm d} \tau}x_1+ \frac{\beta_3 k(\tau) g(x_1,\eta)}{\omega^2}  
=  0, \nonumber
\end{eqnarray}
where most  of  parameters have already been  defined in Eqs. \ref{GLe3} and \ref{GLe4}.
The rest of them  are easily 
derived from Eqs. 
(\ref{GLe8}-\ref{GLe9}), as
\begin{eqnarray}
        \beta_1 & = & 1 / I_2, \nonumber \\
        \beta_2 & = & 1/I_2 + 1/I_3, \nonumber  \\
        \beta_3 & = & (r_2^2/I_2) / \left[ r_1^2/I_1 + r_2^2/I_2 \right], \label{GLe10} \\
             0 & = & r_2M_2/I_2 - r_2M_3/I_3. \nonumber
\end{eqnarray}

\begin{figure}
\vspace{1cm}
\hspace{-0.7cm}
\includegraphics[scale=0.4,angle=-90]{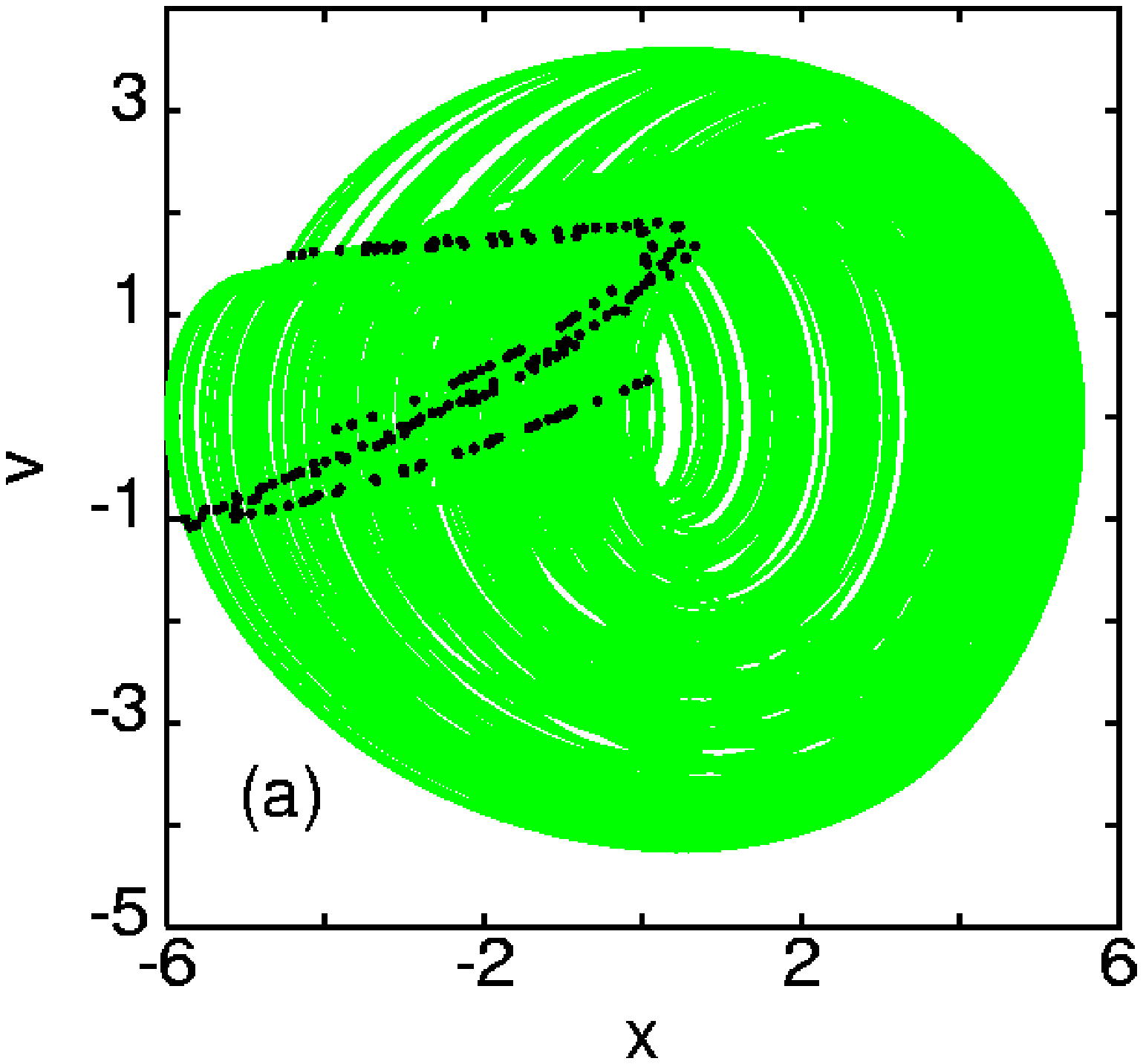} \hspace{-0.8cm}
\includegraphics[scale=0.4,angle=-90]{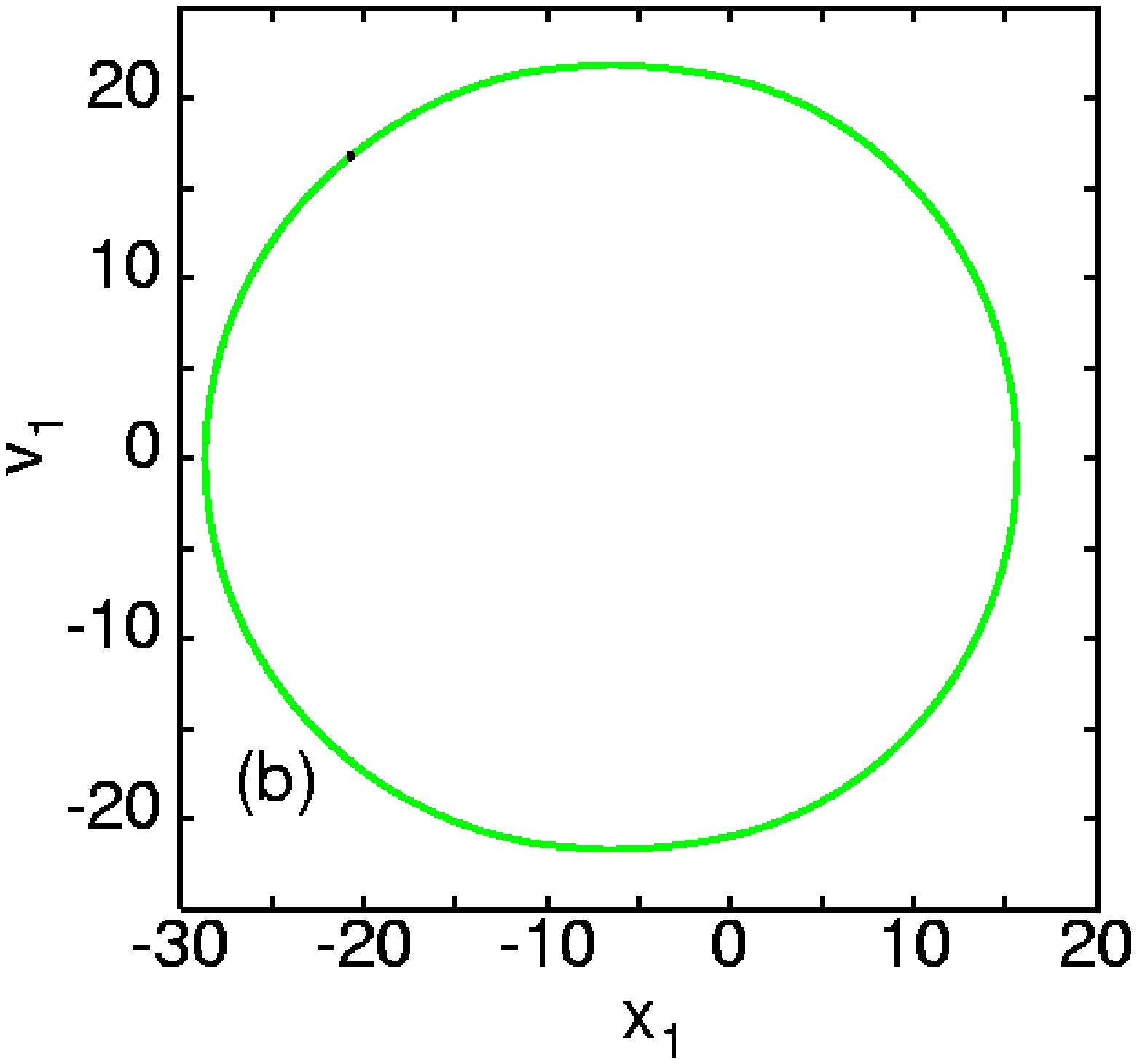}

\hspace{-0.7cm}
\includegraphics[scale=0.4,angle=-90]{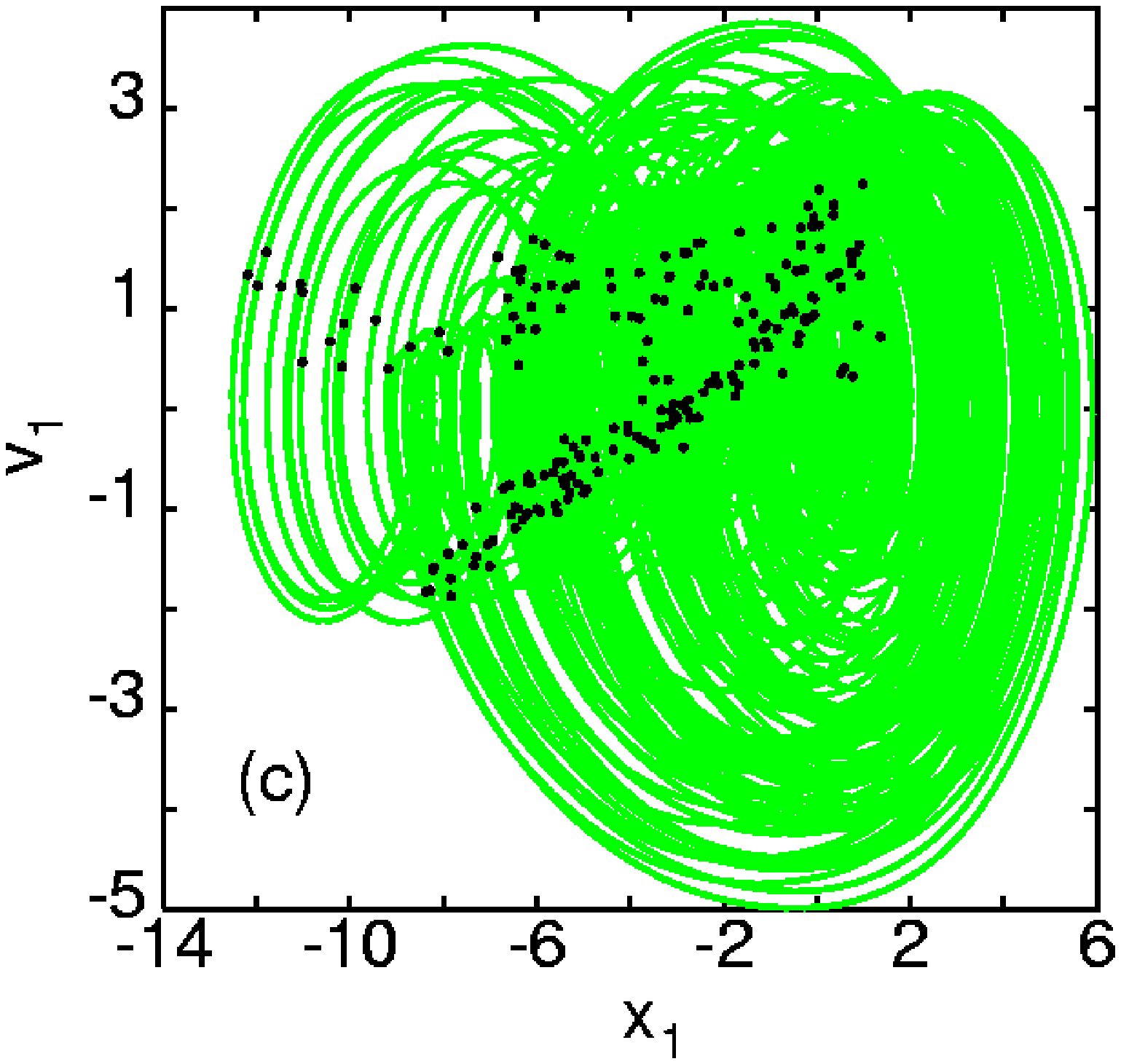} \hspace{-0.8cm}
\includegraphics[scale=0.4,angle=-90]{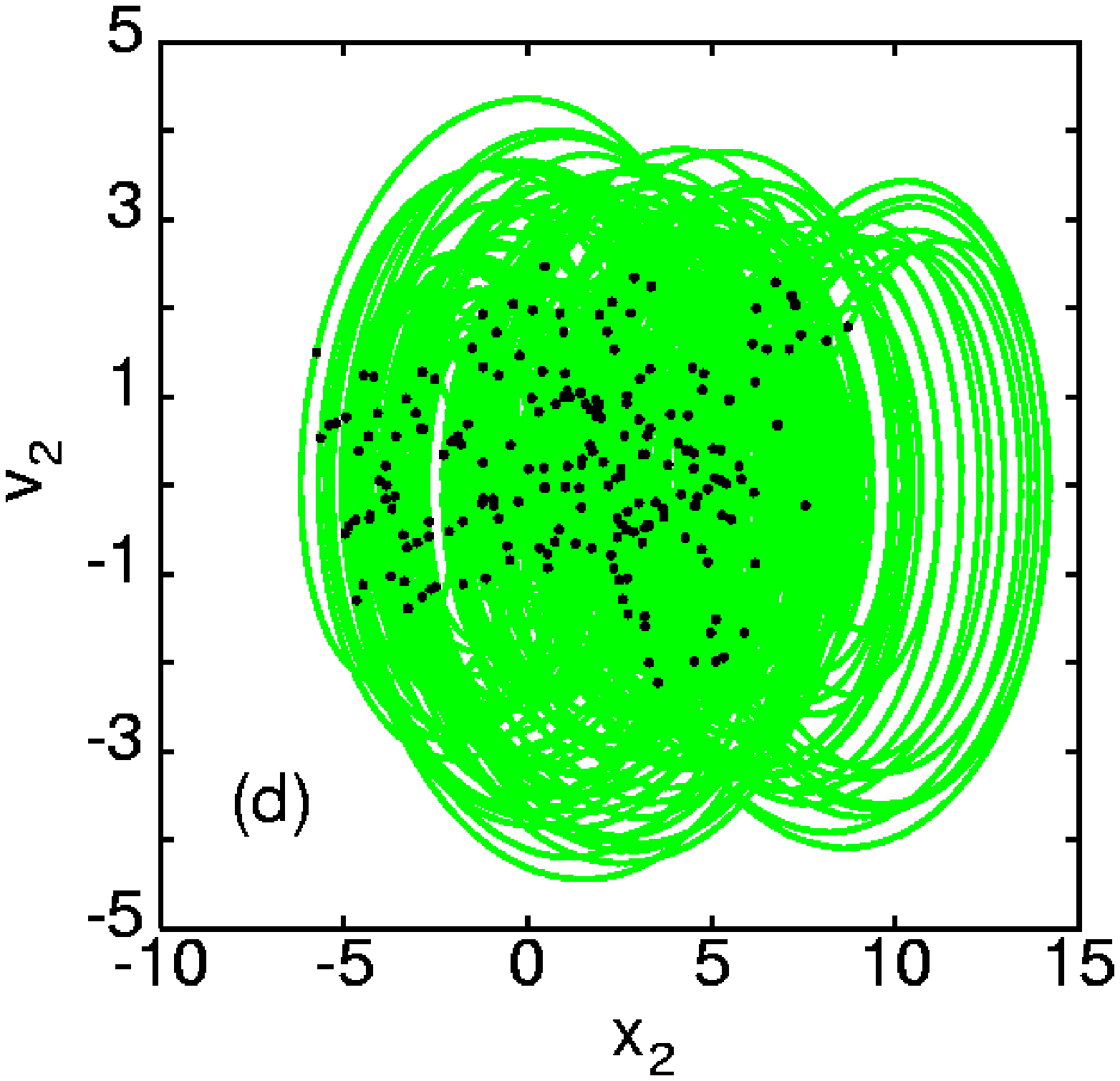}
\caption{\label{GLf7}
 Phase portraits (gray lines) and Poincare maps (black points) for $\omega=1.5$ and $k_S=0$ (a), 
$k_S=1.0$ (b) and $k_S=0.1$ (c-d). The initial conditions:
 $x_{10}=-2.0$,
$v_{10}=\dot{x}_{10}=-0.5$ and $x_{20}=0$ $v_{20}=\dot{x}_{20}=0$.}
\vspace{0cm}
\end{figure}

The simulation results of the above system (Eq. \ref{GLe9}, Fig. \ref{GLf5}) are presented 
in Fig. \ref{GLf6} and \ref{GLf7}. First of all we have repeated the calculations of the amplitude and 
Lyapunov exponents. The results for the amplitude $A$ with various couplings ($k_S$)  are 
summarized  in Fig. \ref{GLf6}a. Curves '1','2','3' correspond to for $k_S=0$, 0.1 and 1.0, respectively.
One can see that a suitable change to $k_S$ may lead to the lowering of the amplitude. 
This effect is more efficient 
in the case of $k_S=1.0$. Comparing the corresponding values of the maximal Lyapunov exponents one  can  
investigate the chaotic nature of the vibrations (Fig. \ref{GLf6}b). Namely, the  case 
$k_S=0.1$ (the curve '2') does not reduce the original chaoticity in region  
around $\omega \approx 1.0$ and  $\omega \in 
[1.24,1.62]$ (the curve '1') while $k_S=1.0$ is large enough to tame the chaos in these regions as well 
as to make the vibration amplitude smaller.   

It should be noted that for $k_S \neq 0$ there are four nonzero exponents to be examined. The 
positive value of the 
largest one $\lambda_1$, is presented in Fig.  
\ref{GLf6}b. It detects frequency regions of chaotic vibrations. But it is also interesting to see what has 
happened to the other 
exponents, especially for our multidimensional system. They were also calculated and are plotted for comparison in Figs.
\ref{GLf6}c-d. For a small coupling stiffness, $k_S=0.1$, we have the surprising result that
 most of the chaotic 
regions are characterized by two positive exponents (Fig. \ref{GLf6}c), which is a signal that the system 
is  truly hyperchaotic \cite{Kap94,War99,War00b}. In this case two initially close  trajectories escape 
exponentially in two different 
directions. The results of calculations with larger $k_S$ ($k_S=1.0$) are different (Fig. \ref{GLf6}), 
where regions of chaotic motion with only one positive Lyapunov exponent were found, which signals 
typical chaotic behaviour. To clarify this point we plot the phase portrait and Poincare maps for the chosen
frequency $\omega=1.5$. Figure \ref{GLf7}a shows the portrait plane and the Poincare map of chaotic 
attractor 
for $k_S=0$
while in Fig. \ref{GLf7}b, for $k_S=1.0$,
  the corresponding attractor is regular, synchronized with the excitation frequency $\omega$. Figures 
\ref{GLf7}c and d show the  phase portrait and the Poincare map of the hyperchaotic attractor.

\section{Conclusions}
This paper has examined the effect of adding an additional small resonance excitation term or an 
additional degree of freedom to a simple
model of gear vibration (Figs. \ref{GLf1} and \ref{GLf2}). We have shown that the resonance term 
can successively reduce chaoticity in the dynamical system (Fig. \ref{GLf3}b), although  
the vibration 
amplitude increases to a larger value than that for the basic system. We also noted several 
sudden jumps in  
the vibration amplitude value (Fig. \ref{GLf3}a) which could destroy a real gearbox. The jumps 
are presumably caused by the 
creation of a larger number of attractors in the presence of the resonance 
term (Fig. \ref{GLf4}).  The extra 
degree of freedom  (Fig. \ref{GLf5}), which may represent a flexible shaft or 
a vibration neutraliser, also has a considerable effect on the dynamics. 
We noted that the suitable use of the coupling value, shaft stiffness $k_S$, can lead to regular
attractors ($k_S=1.0$) and, simultaneously, to a reduction in the vibration amplitude (Fig. 
\ref{GLf6}a-b). Interestingly we have found that for a small value of the coupling stiffness,
$k_S=0.1$, two
of the four Lyapunov exponents were positive, leading to hyperchaos phenomenon. The study of  
the hyperchaotic 
attactor will be reported in a separate article. 
Thus the coupling $k_S$ is a proper bifurcation parameter giving a wide 
range of system behaviour.  The correct choice of this stiffness value can also be used to control the system vibrations.

\renewcommand\bibname{References}

\end{document}